\documentclass[runningheads]{llncs}
\usepackage[T1]{fontenc}
\usepackage{makecell}
\usepackage{siunitx}

\usepackage{tabularx}
\usepackage{booktabs}
\usepackage{graphicx}
\usepackage{subfigure}
\usepackage{cite}
\usepackage{multirow}
\usepackage{xcolor}
\usepackage{amsmath,amsfonts,amssymb}

\usepackage{marvosym}
\usepackage{pifont} 
\usepackage{mathtools}

\usepackage{color}

\usepackage{soul}

\usepackage{hyperref}
\let\svthefootnote\thefootnote
\newcommand\freefootnote[1]{%
  \let\thefootnote\relax%
  \footnotetext{#1}%
  \let\thefootnote\svthefootnote%
}

\usepackage{perpage} 
\MakePerPage{footnote} 

\begin{document}
\title{Adaptive  Spatial Transcriptomics Interpolation  via Cross-modal Cross-slice Modeling}
\titlerunning{Arbitrary Slice Spatial Transcriptomics Interpolation}
\author{Ningfeng Que\inst{1,2}\textsuperscript{$*$} \and
Xiaofei Wang\inst{3}\textsuperscript{$*$}\and
Jingjing Chen\inst{1,2}\and
Yixuan Jiang\inst{1,2}\and
Chao Li\inst{1,3,4,5}\textsuperscript{(\Letter)}
}
\institute{School of Science and Engineering, University of Dundee, UK \and
College of Medicine and Biological Information Engineering, Northeastern University, China \and
Department of Clinical Neurosciences, University of Cambridge, UK \and
School of Medicine, University of Dundee, UK \and
Department of Applied Mathematics and Theoretical Physics, University of Cambridge, UK  \\
correspondence author(\Letter): \email{cl647@cam.ac.uk}
}
\maketitle              

\begin{abstract} Spatial transcriptomics (ST) is a promising technique that characterizes the spatial gene profiling patterns within the tissue context. Comprehensive ST analysis depends on consecutive slices for 3D spatial insights, whereas the missing intermediate tissue sections and high costs limit the practical feasibility of generating multi-slice ST. In this paper, we propose C2-STi, the first attempt for interpolating missing ST slices at arbitrary intermediate positions between adjacent ST slices.
Despite intuitive, effective ST interpolation presents significant challenges, including 1) limited continuity across heterogeneous tissue sections, 2) complex intrinsic correlation across genes, and 3) intricate cellular structures and biological semantics within each tissue section. 
To mitigate these challenges, in C2-STi, we design 1) a distance-aware local structural modulation module to adaptively capture cross-slice deformations and enhance positional correlations between ST slices, 2) 
a pyramid gene co-expression correlation module to capture multi-scale biological associations among genes, and 3) a cross-modal alignment module that integrates the ST-paired hematoxylin and eosin (H\&E)-stained images to filter and align the essential cellular features across ST and H\&E images. 
Extensive experiments on the public dataset demonstrate our superiority over state-of-the-art approaches on both single-slice and multi-slice ST interpolation. Codes are available at \url{https://github.com/XiaofeiWang2018/C2-STi}.

\freefootnote{$*$ Equal contribution.}
\end{abstract}
\section{Introduction}
\label{sec:introduction}

Spatial transcriptomics (ST) has transformed the understanding of tissue function by linking gene expression with spatial context \cite{moses2022museum,tian2023expanding,wang2023multi}. However, in real-world scenarios, tissue sections are often lost or discarded during pre-testing, staining, or quality control, making continuous tissue profiling challenging for current ST platforms \cite{qiu2024spatiotemporal}. Additionally, the high cost of repeated experiments to generate consecutive ST slices limits continuous ST analysis, especially for methods relying on tissue continuity, causing distortions in 3D reconstruction \cite{zhao2024highdimensionalbayesianmodeldiseasespecific}.
This highlights the need for computational methods to overcome the challenges.

Existing approaches focus on super-resolving or imputing regional gene expression within a single slice \cite{wang2024cross}, while few address interpolating the missing ST slices.
A recent study \cite{Monjo2022} proposed a semi-supervised method that first interpolates a missing H\&E slice and then used it to predict the ST. Despite success, 
 the two-step approach may cause bias in leveraging the in-depth spatial information of tissue. In particular, it can only interpolate one single slice at the middle position, assuming the missing tissue is spatially smooth, thus impractical for real-world scenarios that require multiple slices interpolations and reflect the tissue heterogeneity.
 Therefore, it is necessary to develop methods for effectively interpolating ST slices at arbitrary locations between available slices.

In other medical domains, approaches have been proposed to capture the dynamic changes between adjacent slices for estimating the deformation of under-interpolated slices, thereby generating high-quality interpolation. For instance, Frakes \textit{et al.} \cite{4359065} employed a multi-resolution optical flow estimation approach to interpolate 3D anisotropic MRI. However, this method lacks a coarse-to-fine process, making it susceptible to errors for low-quality input. Similarly, Li \textit{et al.} \cite{LI2025126602} developed an interpolation network leveraging joint space-frequency feature extraction and cross-view fusion to achieve CT  reconstruction. However, the alignment errors between different views can introduce artefacts during fusion.

Challenges remain in interpolating the missing ST slices. \textbf{Firstly,} ST slices are often affected by factors such as tissue heterogeneity and experimental procedures, resulting in limited continuity and high deformation between slices. However, traditional interpolation algorithms such as bilinear methods assume strong continuity and fail to capture the nonlinear deformations in tissue structure \cite{li2024vtcnet}. To generate arbitrary ST slices, tailored methods are demanded for learning deformation features at given positions. 
\textbf{Secondly}, ST maps contain pixel-wise expression of thousands of genes, with sparse expression in most genes \cite{Liang2024}. However, existing methods are not tailored to model the spatial gene co-expression in ST maps.
\textbf{Lastly}, biological systems are complex with not only rich gene expression information but also fine cellular structures and biological semantics. Existing methods fail to model gene expression within the context of tissue structures, often resulting in blurred cell boundaries and distorted interpolation. 

To address these challenges, we propose a bespoke framework for ST interpolation at arbitrary positions via cross-modal cross-slice modeling (C2-STi). \textbf{Firstly,} we propose a distance-aware local structural modulation (DLSM) module that leverages distance perception and local tissue structure to  adaptively capture
cross-slice deformations between ST slices, promising to achieve high-quality interpolation at arbitrary positions.
\textbf{Secondly}, we emphasize capturing multi-scale deformation information across slices by modelling gene co-expression. Specifically, We design a pyramid gene co-expression correlation module that integrates a pyramid encoder, a multi-gene co-expression graph (MGC-Graph), and a multi-scale decoder, capturing deformation information between slices at various scales while reflecting gene co-expression relationships.
\textbf{Lastly}, we propose a novel cross-modal alignment module that leverages gated attention weight to integrate H\&E and ST features for preserving key cellular information \cite{zhang2024knowledge}. To our best knowledge, this is the first algorithm to interpolate arbitrary number of missing sections between adjacent ST slices. 
Experiments demonstrate that our method outperforms other state-of-the-art (SOTA) methods.

\begin{figure}[t!]
    \centering
    \includegraphics[width=\textwidth]{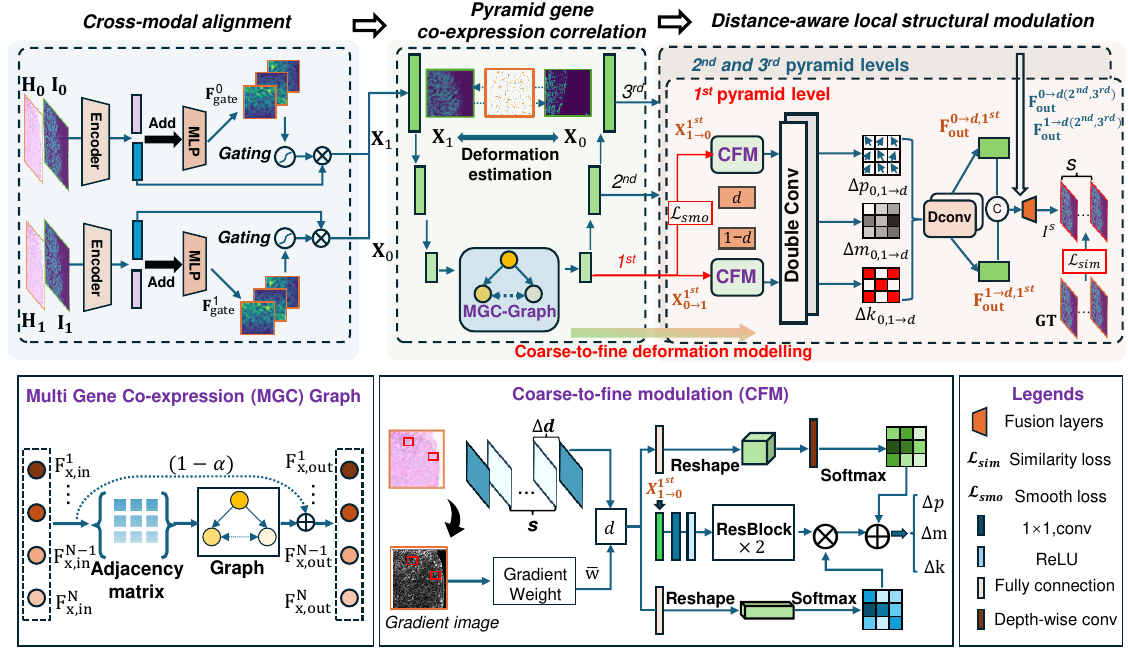} 
    \caption{Framework of the C2-STi, including three modules of cross-modal alignment, pyramid gene co-expression correlation and distance-aware local structural modulation.
    }
    \label{fig:slicework} 
\end{figure}
\section{Methodology}

Fig. \ref{fig:slicework} illustrates the proposed C2-STi. Our model achieves St interpolation by adaptively interpolating the missing slices through cross-modal cross-section modeling, leveraging H\&E images and ST maps from adjacent slices. 
C2-STi takes as input two adjacent ST slices of $N$ genes, $\mathbf{I}_0$ and $\mathbf{I}_1$, along with their corresponding H\&E images $\mathbf{H}_0$ and $\mathbf{H}_1$. It generates $s$ imputed intermediate ST slices, $\{\mathbf{I}^{i}_{\text{out}}\}_{i=1}^s$.
In general, three key modules are proposed in C2-STi, including \textbf{1)} a  cross-modal alignment module (Section \ref{section_cross})  to filter and match the essential cellular features across modalities, \textbf{2)} a pyramid gene co-expression correlation module (Section \ref{section_graph}) for capturing multi-scale associations among genes, and \textbf{3)} a distance-aware local structural modulation module (Section \ref{section_Distance}) to model the the spatial relationships and preserve context consistency across ST slices.

\subsection{Cross-modal alignment between H\&E and ST}\label{section_cross}

Previous studies show a strong correlation between histological features and gene expression \cite{Zhang2024}, with cellular details in H\&E-stained images supporting ST interpolation. Leveraging commonly available ST-paired H\&E images could thus enhance interpolation effectiveness.
However, H\&E and ST images capture distinct information: H\&E highlights cellular patterns, while ST maps focus on gene expression. To bridge the differences and extract cellular features for ST interpolation, we propose a cross-modal alignment module, as shown in Fig. \ref{fig:slicework}.

The cross-modal alignment module takes two adjacent ST slices $\mathbf{I}_0$ and $\mathbf{I}_1$ as input, along with their paired H\&E slices $\mathbf{H}_0$ and $\mathbf{H}_1$. For each of the ST-H\&E pair, we first use the  pre-trained ResNet-50 \cite{10.1145/3225058.3225069} to extract features $\mathbf{M}_\mathbf{h} \in \mathbb{R}^{C_h \times H \times W}$ (with $C_h$, $H$ and $W$ denoting the channel number, height and width of the features) and $\mathbf{M}_\mathbf{s} \in \mathbb{R}^{C_s \times H \times W}$ for the H\&E and ST image, respectively. Then, to mitigate the impact of inherent noise from each modality on the overall fusion, we further conduct gated-attention-based cross-modal alignment on the concatenated multimodal features $\mathbf{F}_{\text{cat}} = [\mathbf{M}_\mathbf{h}; \mathbf{M}_\mathbf{s}] \in \mathbb{R}^{(C_s+C_h) \times H \times W}$. Specifically, we used two $1 \times 1$ convolutional layers coupled with activations to generate the attention weights $\mathbf{G}$ for retaining useful ST features with rich cellular patterns. Formally, $\mathbf{G}$ is defined as: 
\begin{equation}
G = \text{Sigmoid}\left(\text{Conv}_{1 \times 1}(\mathbf{F}_{\text{gate}})\right), \text{where} \;\mathbf{F}_{\text{gate}} = \text{ReLU}\left(\text{Conv}_{1 \times 1}(\mathbf{F}_{\text{cat}})\right)
\end{equation}
Finally, we obtained the cross-modal aligned features \( \mathbf{X} = \mathbf{M}_\mathbf{s} \odot {\mathbf{G}}\), where \( \odot \) represents element-wise multiplication, for each of the two ST maps.

\subsection{Pyramid gene co-expression correlation} 
\label{section_graph}

As shown above, effective ST slice interpolation requires  both tissue-level and cellular-level deformations while considering gene-specific variations . To address this, we propose the pyramid gene co-expression correlation module, shown in Fig. \ref{fig:slicework}, which integrates a pyramid encoder-decoder structure with a multi-gene co-expression graph (MGC-Graph). Details follow the processing flow below.

\noindent\textbf{Pyramid feature  encoder.} 
For each cross-modal fused feature $\mathbf{X}_0$ and $\mathbf{X}_1$ from adjacent ST slices, we design a weight-shared extractor to capture contextual information using pyramid features. At each pyramid level $L \in \{1, 2, 3\}$, four convolutional layers perform spatial downsampling, producing the final multi-scale features $ \mathbf{C}^{L}_0, \mathbf{C}^{L}_1 $, each with the feature channel of $C$.

\noindent\textbf{MGC-Graph.} To capture the co-expression relationships among genes in adjacent ST slices, we propose MGC-Graph, denoted as $\mathbf{g} = (\mathcal{V}, \mathcal{E})$, where $\mathcal{V}$ represents nodes (genes)  and $\mathcal{E}$ represents edges based on co-expression intensity. 

Specifically, given the output features of the pyramid feature  encoder \( \{ \mathbf{C}^{L}_0, \mathbf{C}^{L}_1 \} \) as input nodes, we construct a multi-gene co-expression based correlation matrix $\mathbf{A} \in \mathbb{R}^{N\times N}$ to reflect the relationships among each node feature, with a weight matrix $\mathbf{W}_{n} \in \mathbb{R}^{C\times C}$ to update the value of $\mathbf{C}^{L}_0$ and $\mathbf{C}^{L}_1$. Formally, the correlation matrix $\mathbf{A}$ can be defined as:
\begin{equation}
    \mathbf{A} = \left[ \rho_{ik} \right]_{i,k=1}^{N}, \text{where} \; 
\rho_{ik} = \tilde{\mathbf{E}}_i \cdot \tilde{\mathbf{E}}_k, \text{and} \;  
\tilde{\mathbf{E}}_i = \frac{\mathbf{E}_i - \bar{\mathbf{E}}_i}{\|\mathbf{E}_i - \bar{\mathbf{E}}_i\|_2^2}.
\end{equation}
where $\cdot$ denotes the calculation of Pearson correlation coefficient (PCC), and \(\bar{\mathbf{E}}_i\) represents the average expression intensity of $i$-th gene across ST slices. Based on the correlation matrix $\mathbf{A}$, the output nodes $\mathbf{F}^{L}_0$ of $\mathbf{C}^{L}_0$ are formulated by a single graph convolutional network(GCN) layer as
\begin{equation}\label{eq1}
\mathbf{F}^{L}_0 = \lambda\,\mathbf{C}^{L}_{0, graph} + (1 - \lambda)\,\mathbf{C}^{L}_0,\text{where}\; 
\mathbf{C}^{L}_{0, graph} = \delta\bigl(\mathbf{P}\,\mathbf{C}^{L}_0\,W_n\bigr).
\end{equation}
where $\mathcal{\delta} (\cdot)$ is an activation function, and $\lambda$ is a graph balancing hyper-parameter  Similarly, the output nodes $\mathbf{F}^{L}_1$ of $\mathbf{C}^{L}_1$ are obtained using the same GCN layer.

\noindent\textbf{Multi deformation estimation decoder.} After extracting  hierarchical information, we add a separate decoder $D^{(L)}$ for the feature maps $\mathbf{F}^{L}_0$ and $\mathbf{F}^{L}_1$ of multiple scales. Specifically,  the coarse deformation features between $\mathbf{F}_0$ and $\mathbf{F}_1$  are processed as follows:
$\{ \mathbf{F}_{0 \to 1}^{n}, \mathbf{F}_{1 \to 0}^{n} \}_{n=1}^{L}\ =D^{(L)} (\mathbf{F}^{L}_0, \mathbf{F}^{L}_1)$.

\subsection{ Distance-aware local structural modulation across ST slices}
\label{section_Distance}

To enhance structural consistency across input ST slices and enable interpolation at arbitrary positions, we propose the DLSM module, shown in Fig. \ref{fig:slicework}. It refines coarse deformation features $\{ \mathbf{F}_{0 \to 1}^{n}, \mathbf{F}_{1 \to 0}^{n} \}_{n=1}^{L}$ based on slice positions. The module comprises three key components: cross-section distance modeling, coarse-to-fine modulation, and deformable convolutional fusion layers.

\noindent\textbf{Cross-section distance modeling.} 
As shown in Fig. \ref{fig:slicework}, we model the distance between adjacent ST slices to to quantify biological structure similarity and enable effective interpolation. 
Given $s$ under-interpolated slices and interval  $\Delta d$  interval between two adjacent ST slices, the distance between any two slices $\mathbf{I}_{i}$ and $\mathbf{I}_{j}$, across the entire set of $s$ slices can then be expressed as $d_{i,j}=(s+1)\Delta d$.

Next, the cross-slice distance can be further refined via weighted on the amplitude of the gradient map\footnote{The gradient map can capture the structural information of an image \cite{liu2011image}.} of the H\&E images  to emphasize regions with significant structural changes.  Specifically, the Sobel operator \cite{kanopoulos1988design} is used to calculate the gradient-based weight of $i$-th slice as $\overline{w}_i = \text{Norm}(1 + \alpha |\nabla \mathbf{H}_i|)$, where $\alpha$ is a gradient scaling parameter.  Thus, the position set of the under-interpolated slices is obtained as $\mathbf{P}=\{ \mathrm{p}_i \;|\;\mathrm{p}_i=\frac{i}{s}\cdot \overline{w}_i \in (0,1), \; i  \;\text{values from }  1 \;\text{to} \; s-1\}$.

\noindent\textbf{Coarse-to-fine modulation.} Based on the position set $\mathbf{P}$, we  design a coarse-to-fine modulation to generate multiple deformation features at different positions, adaptively adjusting the coarse deformation features $\{ \mathbf{F}_{0 \to 1}^{n}, \mathbf{F}_{1 \to 0}^{n} \}_{n=1}^{L}$. The forward feature $\mathbf{F}_{0 \to 1}$ is modulated by $\{\mathrm{p}_i\}_{i=1}^{s-1}$, while the backward feature $\mathbf{F}_{1 \to 0}$ is modulated by $\{1-\mathrm{p}_i\}_{i=1}^{s-1}$  for 
symmetry. The process is detailed below.

First, we apply global average pooling to $\mathbf{F}_{0 \to 1}$ to obtain a statistical vector \( \bar{\mathbf{f}} \), while the positions $\mathbf{P}=\{\mathrm{p}_i\}_{i=1}^{s-1}$ are mapped to the embedding vector \( \mathbf{e}_p = \phi(\mathbf{P}) \in \mathbb{R}^d \). Concatenating \( \bar{\mathbf{f}} \) and \( \mathbf{e}_p \) to obtain  global context and positional information. Further,  the concatenated features are fed into fully-connected layers to generate channel weight \( \mathbf{w}_c \), used for obtaining the channel-adaptive feature \( \mathbf{F}_{\text{ca}} = \mathbf{w}_c \odot \mathbf{F}_{0 \to 1} \), where $\odot$ denotes Kronecker product. 
Moreover, to extract local  details, a \( 3\times3 \) convolutional  is applied to $\mathbf{F}_{0 \to 1}$ to obtain local structural features \( \mathbf{F}_{\text{loc}}\). The positions $\mathbf{P}=\{\mathrm{p}_i\}_{i=1}^{s-1}$ are mapped by another embedding function to  \( \mathbf{e}'_p = \psi(\mathbf{P}) \in \mathbb{R}^{d'} \) and  expanded into a tensor \( \mathbf{E}_p \in \mathbb{R}^{d' \times H \times W} \). 

Next, we  use  \( 1\times1 \) convolution and  multilayer perceptron (MLP) to obtain the fused feature \( \mathbf{F}_{\text{fused}}\). Then, taking \(\mathbf{E}'_p\), a small MLP is utilized to generate a adjustable  kernel \( k \) applied to \( \mathbf{F}_{\text{fused}} \) through depth-wise  convolution, further generating the spatial-adaptive feature \( \mathbf{F}_{\text{sp}} \). Further, the output of the channel branch \( \mathbf{F}_{\text{ca}} \) and the spatial branch \( \mathbf{F}_{\text{sp}} \) are combined using preset weights \(\beta\) and \(\gamma\) to obtain the final fully modulated deformable feature \( \mathbf{F}_{\text{out}} = \beta \mathbf{F}_{\text{ca}} + \gamma \mathbf{F}_{\text{sp}} \).

Finally, three estimators  predict adjustments for $\mathbf{F}_{\text{out}}^{0 \to 1}$ with its position-related kernel $\Delta k_{0\to d}$, offsets $\Delta p_{0\to d}$, and masks $\Delta m_{0\to d}$ as follows:

\begin{equation}
\left\{
\begin{array}{lr}
    \Delta k_{0\to d} \;= \text{KernelEstimator}(\mathbf{F}_{\text{out}}) \in \mathbb{R}^{C \times k_h \times k_w}, \\
    \Delta p_{0\to d} \;= \text{OffsetEstimator}(\mathbf{F}_{\text{out}}) \in \mathbb{R}^{\text{offset\_dim}}, \\
    \Delta m_{0\to d} \;= \text{MaskEstimator}(\mathbf{F}_{\text{out}}) \in \mathbb{R}^{\text{mask\_dim}}.
    \end{array}
    \right.
\end{equation}

Similarly, the fully modulated deformable feature $\mathbf{F}_{\text{out}}^{1 \to 0}$ for $\mathbf{F}_{1 \to 0}$ can also be obtained via the above process.

\noindent\textbf{Deformable convolution fusion layers:} Based on the learned offsets, masks, and kernels, we further use the deformable convolution \cite{shi2021videoframeinterpolationgeneralized}, denoted as \(\mathcal{F}_{\text{DConv}}\), to  generate the intermediate interpolation feature maps:   

\begin{equation}
\left\{
\begin{array}{lr}
    \mathbf{F}_{\text{out}}^{{0 \to d}} &= \mathcal{F}_{\text{DConv}}(\mathbf{F}_{0 \to 1}, \Delta p_{0 \to d}, \Delta m_{0 \to d}, \Delta k_{0 \to d}), \\
    \mathbf{F}_{\text{out}}^{{1 \to d}} &= \mathcal{F}_{\text{DConv}}(\mathbf{F}_{1 \to 0}, \Delta p_{1 \to d}, \Delta m_{1 \to d}, \Delta k_{1 \to d}).
        \end{array}
    \right.
\end{equation}

At each pyramid level, \(\mathbf{F}_{\text{out}}^{{0 \to d}}\) and \(\mathbf{F}_{\text{out}}^{{1 \to d}}\) are fused  and  upsampled with a pixel-shuffle layer to obtain the imputed $s$ intermediate ST slices \( \{\mathbf{I}^{i}_{\text{out}}\}_{i=1}^s\).

\subsection{Loss function}
We use two loss functions to align the interpolated slice $\mathbf{I}^{s}$ and ground truth $\mathbf{I}^{gt}$:
\(\mathcal{L}_{\text{sim}} = \sum_{s} \| \mathbf{I}^{gt} - \mathbf{I}^{s} \|_1\) measuring the pixel-wise discrepancy, and
\(\mathcal{L}_{\text{smo}} = \|\nabla F_{0 \to 1}\|_1 + \|\nabla F_{1 \to 0}\|_1\) regularizing pixel variations in the interpolated slice. 
The final loss function is expressed as:

\(\mathcal{L} = \lambda_{\text{sim}} \mathcal{L}_{\text{sim}} + \lambda_{\text{smo}} \mathcal{L}_{\text{smo}}\), 
where $\lambda_{\text{sim}}=1$ and $\lambda_{\text{smo}}=1$ control the balance between accuracy and smoothness.

\section{Experiments \& Results}

\noindent\textbf{Dataset.} we use the largest 3D H\&E-ST dataset, i.e, HNSCC \cite{Schott2024OpenST}, which includes 1,216 aligned H\&E-ST image pairs from 19 consecutive slices spanning 350 mm. The patches are  with the size of 512 × 512 and splitted into the training (852), validation (100), and test (264) sets. For the single-slice interpolation task, for each of the three consecutive patches, we use the 1\textit{st} and 3\textit{rd} slices as input, and the 2\textit{nd} slice as the targeted slice for interpolation.
Similarly, for $N$-slice interpolation tasks, we further construct the input-output pair using each of the $N$+2 consecutive patches of the whole dataset.

\noindent\textbf{Implementation details.} Our method is optimized using AdamW \cite{loshchilov2017decoupled} and implemented with PyTorch. We trained our model for 40 epochs
on seven NVIDIA RTX V100 32 GB GPUs,  with a weight decay of $10^{-4}$. The initial learning rate is set to $10^{-4}$, which is decreased to $10^{-6}$ using cosine annealing. The batch size for each training step is set to 6.

\begin{figure}[t!]
    \centering
    \includegraphics[width=.8\textwidth]{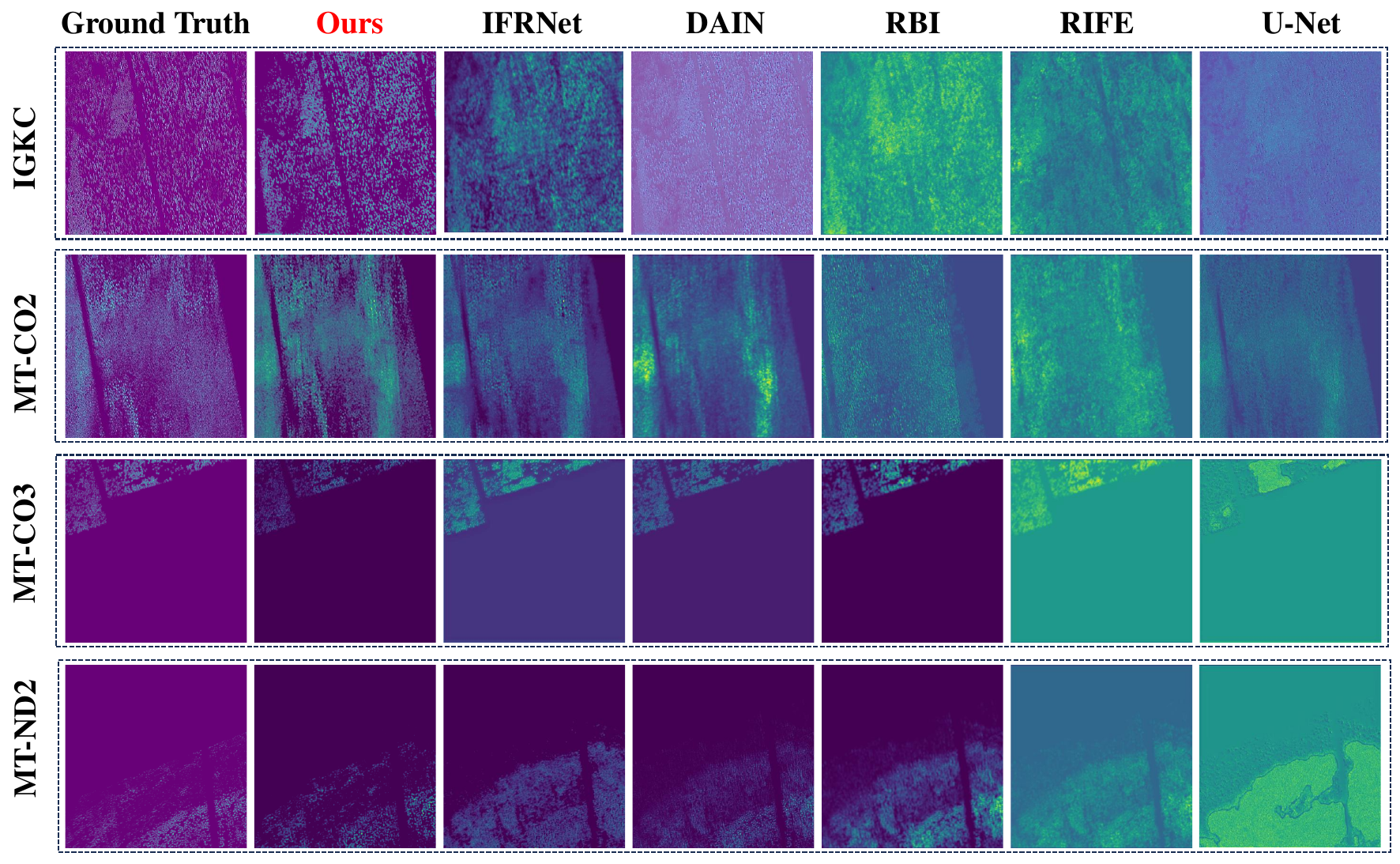} 
    \caption{Visual comparisons  for single slice interpolation result. Note that IGKC, MT-CO2, MT-CO3, and MT-ND2 denote different genes.}
    \label{fig:experiment} 
\end{figure}

\subsection{Performance evaluation}

\noindent\textbf{Single slice interpolation.} We compared our model with five other SOTA methods, i.e., IFRNet \cite{kong2022ifrnetintermediatefeaturerefine}, RBI \cite{1309715}, DAIN \cite{bao2019depthawarevideoframeinterpolation}, RIFE \cite{huang2022realtimeintermediateflowestimation} and U-Net \cite{park2021unetbasedglowopticalflowfree}. Of note, IFRNet, DAIN and RIFE are SOTA  video frame interpolation approaches, while RBI is MR image interpolation method, and U-Net serves as the baseline.    
The left panel of Table \ref{tab:table1} shows that C2-STi outperforms all competing methods in single-slice interpolation. Compared to the other five methods, our method improves the peak signal-to-noise ratio (PSNR), structural similarity index measure (SSIM), and PCC by at least 6.0\%, 6.6\%, and 29.0\%, respectively, while also advancing the root mean square error (RMSE) by at least 21.4\%. These results demonstrate that our method effectively recovers slice information, enhances image quality, and preserves structural consistency.

\begin{table}[t]
\scriptsize
\centering
\caption{Comparisons of different interpolation methods for single slice interpolation.}
\label{tab:table1}
\begin{tabular}{cccccccccccccccccc}
    \toprule
    \multirow{1}*{Method} & \multicolumn{11}{c}{Interpolation network}   &\multicolumn{1}{p{0.3em}}{} & \multicolumn{5}{c}{Ablation \textit{(w/o)}} \\
    \specialrule{0em}{1pt}{1pt}
    \cline{2-12}  
    \cline{14-18}
    \specialrule{0em}{2pt}{2pt}
      & RIFE  &\multicolumn{1}{p{0.1em}}{} &RBI  &\multicolumn{1}{p{0.0em}}{} & DAIN  &\multicolumn{1}{p{0.2em}}{} & IFRNet &\multicolumn{1}{p{0.2em}}{} & Unet &\multicolumn{1}{p{0.2em}}{} & Ours &\multicolumn{1}{p{0.2em}}{}& Cross-modal &\multicolumn{1}{p{0.2em}}{}& MGC-Graph &\multicolumn{1}{p{.2em}}{} & DLSM\\

    \midrule

PSNR$\uparrow$ & 41.02 &\multicolumn{1}{p{0em}}{} & 41.47 &\multicolumn{1}{p{0em}}{} & 44.48 &\multicolumn{1}{p{0em}}{}& 51.07 &\multicolumn{1}{p{0em}}{}& 40.21 &\multicolumn{1}{p{0em}}{}&\textbf{54.11}&\multicolumn{1}{p{0em}}{} & 48.72 &\multicolumn{1}{p{0em}}{}& 50.15 &\multicolumn{1}{p{0em}}{} & 48.80  \\

SSIM$\uparrow$ & 0.76 &\multicolumn{1}{p{0em}}{}& 0.71 &\multicolumn{1}{p{0em}}{} & 0.76 &\multicolumn{1}{p{0em}}{} & 0.73&\multicolumn{1}{p{0em}}{} & 0.61&\multicolumn{1}{p{0em}}{} &\textbf{0.81}&\multicolumn{1}{p{0em}}{} & 0.77 &\multicolumn{1}{p{0em}}{} & 0.78 &\multicolumn{1}{p{0em}}{}& 0.77  \\

PCC$\uparrow$ & 0.17 &\multicolumn{1}{p{0em}}{}& 0.23 &\multicolumn{1}{p{0em}}{} & 0.33 &\multicolumn{1}{p{0em}}{} & 0.38 &\multicolumn{1}{p{0em}}{} & 0.15&\multicolumn{1}{p{0em}}{}& \textbf{0.49}&\multicolumn{1}{p{0em}}{} & 0.42 &\multicolumn{1}{p{0em}}{} & 0.47 &\multicolumn{1}{p{0em}}{}& 0.47  \\

RMSE$\downarrow$ & 0.16 &\multicolumn{1}{p{0em}}{}& 0.19 &\multicolumn{1}{p{0em}}{} & 0.15 &\multicolumn{1}{p{0em}}{} & 0.14&\multicolumn{1}{p{0em}}{} &  0.23 &\multicolumn{1}{p{0em}}{}& \textbf{0.11}&\multicolumn{1}{p{0em}}{} & 0.13&\multicolumn{1}{p{0em}}{} & 0.12&\multicolumn{1}{p{0em}}{}& 0.12 \\

\bottomrule
\end{tabular}
\end{table}

\noindent\textbf{Multi-slice interpolation.} The right panel of Table \ref{table2} shows the performance of various interpolation methods under multi-slice settings (2, 3, and 4 slices). Our model consistently outperforms all competitors. Specifically, in the 2-slice scenario, our approach achieves a PSNR of 43.90 dB, which is at least 3.2\% higher than the other methods. Similarly, for the 3-slice case, C2-STi attains the best SSIM of 0.77, representing an improvement of at least 2.5\% over the alternatives. Finally, in the 4-slice setting, our method demonstrates robust performance by achieving a PSNR that is at least 3.1\% greater than those of the compared approaches. These results highlight that our model not only excels in single-slice interpolation but also in handling more complex multi-slice scenarios.

\subsection{Ablation Experiments} 

We assess the contribution of three key components in our model: 1) \textit{w/o} Cross-modal – using only the ST modality; 2) \textit{w/o} MGC-Graph – removing gene co-expression modeling; and 3) \textit{w/o} DLSM – inserting slices at predefined positions without considering structural correlations. The ablation results of single-slice and multi-slice interpolation  tasks are tabulated in Table~\ref{tab:table1} and Table~\ref{table2}, respectively.
As shown in both tables, all three ablated models perform worse than the complete model, suggesting that these components substantially enhance overall performance. Notably, the the \textit{w/o} Cross-modal variant  performs the worst, consistent with our hypothesis that conventional fusion strategies fail to fully leverage complementary multi-modal information.

\begin{table*}[t]
\centering
\caption{Comparisons of different interpolation methods for multi-slices task}
\label{table2}
 \renewcommand{\arraystretch}{1.3} 
 \resizebox{\textwidth}{!}{%
    \begin{tabular}{cc ccccccccccccc}
        \toprule
        \multirow{2}{*}{\textbf{}} & \multirow{2}{*}{\textbf{Method}} & \multirow{2}{*}{\textbf{Position-based}} & \multicolumn{4}{c}{\textbf{2 slices}} & \multicolumn{4}{c}{\textbf{3 slices}} & \multicolumn{4}{c}{\textbf{4 slices}} \\
        \cmidrule(lr){4-7}\cmidrule(lr){8-11}\cmidrule(lr){12-15}
         & & & PSNR & RMSE & PCC& SSIM
           & PSNR & RMSE & PCC & SSIM
           & PSNR & RMSE & PCC & SSIM \\
        \midrule
        & RIFE    & Pre-defined & 41.50 & 0.29 & 0.34 & 0.68 & 38.30 & 0.27 & 0.25 & 0.58 & 39.80 & 0.29 & 0.21 & 0.57 \\
        & DAIN    & Pre-defined & 42.30 & 0.28 & 0.42 & 0.72 & 39.10 & 0.27 & 0.25& 0.64 & 40.50 & 0.27 &0.24 & 0.51 \\
        & IFRNet  & Pre-defined & 40.75 & 0.31 & 0.45& 0.74 & \textbf{42.95} & 0.21 & 0.28 & 0.65 & 39.30 & 0.22 & 0.27 & 0.74 \\
        & RBI     & Adaptive    & 42.80 & 0.25 & 0.17 & 0.61 & 35.50 & 0.30 & 0.15 &0.59  & 33.30 & 0.33 & 0.17 & 0.47 \\
        & Unet    & Pre-defined & 43.10 & 0.27 & 0.15 & 0.59 & 33.00 & 0.28 & 0.14 & 0.55 & 33.20 & 0.36 & 0.13 & 0.44 \\
        \specialrule{1.5pt}{0pt}{0pt}
        & \textit{w/o} Cross-modal  & Adaptive    & 43.30 & 0.18 & 0.43 & 0.72 & 42.60 & 0.20 & 0.28 & 0.72 & 41.80 & 0.19 & 0.34 & 0.70 \\
        & \textit{w/o }MGC-Graph    & Adaptive    & 43.70 & 0.16 & 0.40 & 0.78 & 42.40 &0.22 & 0.26  & 0.72 & 41.90 & 0.19 & 0.35 & 0.69 \\
        & \textit{w/o}  DLSM & Pre-defined & 43.00 & 0.14 & 0.44 & 0.77 & 42.20 & 0.18 & 0.28 & 0.73 & 41.20 & 0.19 & 0.33 & 0.72 \\
        \specialrule{1.5pt}{0pt}{0pt}
         & \textbf{Ours} & Adaptive & \textbf{43.90} & \textbf{0.11} & \textbf{0.46} &  \textbf{0.78}  & 42.77 & \textbf{0.18} & \textbf{0.34} & \textbf{0.77} & \textbf{42.00}& \textbf{0.19}& \textbf{0.35} & \textbf{0.76}\\
        \bottomrule
    \end{tabular}}
\end{table*}

\section{Conclusion}

In this study, we introduced C2-STi, the first work for interpolating missing ST slices at arbitrary intermediate positions. Addressing key challenges in ST interpolation, our approach ensures structural continuity across heterogeneous tissue sections, models  gene correlations, and preserves intricate cellular structures. By integrating a distance-aware local structural modulation module, a pyramid gene co-expression correlation module, and a cross-modal alignment module, C2-STi effectively captures cross-slice deformations, enhances positional correlations, and leverages H\&E images for refinement. Extensive experiments on a public dataset demonstrate its superiority over SOTA methods in both single-slice and multi-slice ST interpolation.

\clearpage
\bibliographystyle{splncs04}
\bibliography{main}

\begin{thebibliography}{10}
\providecommand{\url}[1]{\texttt{#1}}
\providecommand{\urlprefix}{URL }
\providecommand{\doi}[1]{https://doi.org/#1}

\bibitem{bao2019depthawarevideoframeinterpolation}
Bao, W., Lai, W.S., Ma, C., Zhang, X., Gao, Z., Yang, M.H.: Depth-aware video
  frame interpolation (2019), \url{https://arxiv.org/abs/1904.00830}

\bibitem{4359065}
Frakes, D.H., Dasi, L.P., Pekkan, K., Kitajima, H.D., Sundareswaran, K.,
  Yoganathan, A.P., Smith, M.J.T.: A new method for registration-based medical
  image interpolation. IEEE Transactions on Medical Imaging  \textbf{27}(3),
  370--377 (2008). \doi{10.1109/TMI.2007.907324}

\bibitem{huang2022realtimeintermediateflowestimation}
Huang, Z., Zhang, T., Heng, W., Shi, B., Zhou, S.: Real-time intermediate flow
  estimation for video frame interpolation (2022),
  \url{https://arxiv.org/abs/2011.06294}

\bibitem{kanopoulos1988design}
Kanopoulos, N., Vasanthavada, N., Baker, R.L.: Design of an image edge
  detection filter using the sobel operator. IEEE Journal of solid-state
  circuits  \textbf{23}(2),  358--367 (1988)

\bibitem{kong2022ifrnetintermediatefeaturerefine}
Kong, L., Jiang, B., Luo, D., Chu, W., Huang, X., Tai, Y., Wang, C., Yang, J.:
  Ifrnet: Intermediate feature refine network for efficient frame interpolation
  (2022), \url{https://arxiv.org/abs/2205.14620}

\bibitem{li2024vtcnet}
Li, M., Que, N., Zhang, J., Du, P., Dai, Y.: Vtcnet: A feature fusion dl model
  based on cnn and vit for the classification of cervical cells. International
  Journal of Imaging Systems and Technology  \textbf{34}(5),  e23161 (2024)

\bibitem{LI2025126602}
Li, W., Song, H., Ai, D., Shi, J., Fan, J., Xiao, D., Fu, T., Lin, Y., Wu, W.,
  Yang, J.: Sfcli-net: Spatial-frequency collaborative learning interpolation
  network for computed tomography slice synthesis. Expert Systems with
  Applications  \textbf{272},  126602 (2025).
  \doi{https://doi.org/10.1016/j.eswa.2025.126602},
  \url{https://www.sciencedirect.com/science/article/pii/S0957417425002246}

\bibitem{Liang2024}
Liang, Y., Shi, G., Cai, R., Yuan, Y., Xie, Z., Yu, L., Huang, Y., Shi, Q.,
  Wang, L., Li, J., Tang, Z.: Prost: quantitative identification of spatially
  variable genes and domain detection in spatial transcriptomics. Nature
  Communications  \textbf{15}(1), ~600 (January 2024).
  \doi{10.1038/s41467-024-44835-w},
  \url{https://doi.org/10.1038/s41467-024-44835-w}

\bibitem{liu2011image}
Liu, A., Lin, W., Narwaria, M.: Image quality assessment based on gradient
  similarity. IEEE Transactions on Image Processing  \textbf{21}(4),
  1500--1512 (2011)

\bibitem{loshchilov2017decoupled}
Loshchilov, I., Hutter, F.: Decoupled weight decay regularization. arXiv
  preprint arXiv:1711.05101  (2017)

\bibitem{Monjo2022}
Monjo, T., Koido, M., Nagasawa, S., Suzuki, Y., Kamatani, Y.: Efficient
  prediction of a spatial transcriptomics profile better characterizes breast
  cancer tissue sections without costly experimentation. Scientific Reports
  \textbf{12}(1), ~4133 (2022)

\bibitem{moses2022museum}
Moses, L., Pachter, L.: Museum of spatial transcriptomics. Nature methods
  \textbf{19}(5),  534--546 (2022)

\bibitem{park2021unetbasedglowopticalflowfree}
Park, S., Han, D., Kwak, N.: The u-net based glow for optical-flow-free video
  interframe generation (2021), \url{https://arxiv.org/abs/2103.09576}

\bibitem{1309715}
Penney, G., Schnabel, J., Rueckert, D., Viergever, M., Niessen, W.:
  Registration-based interpolation. IEEE Transactions on Medical Imaging
  \textbf{23}(7),  922--926 (2004). \doi{10.1109/TMI.2004.828352}

\bibitem{qiu2024spatiotemporal}
Qiu, X., Zhu, D.Y., Lu, Y., Yao, J., Jing, Z., Min, K.H., Cheng, M., Pan, H.,
  Zuo, L., King, S., et~al.: Spatiotemporal modeling of molecular holograms.
  Cell  \textbf{187}(26),  7351--7373 (2024)

\bibitem{Schott2024OpenST}
Schott, M., León-Periñán, D., Splendiani, E., Strenger, L., Licha, J.R.,
  Pentimalli, T.M., Schallenberg, S., Alles, J., Tagliaferro, S.S.,
  Boltengagen, A., Ehrig, S., Abbiati, S., Dommerich, S., Pagani, M., Ferretti,
  E., Macino, G., Karaiskos, N., Rajewsky, N.: Open-st: High-resolution spatial
  transcriptomics in 3d. Cell  \textbf{187}(15),  3953--3972.e26 (July 2024).
  \doi{10.1016/j.cell.2024.05.055},
  \url{https://doi.org/10.1016/j.cell.2024.05.055}

\bibitem{shi2021videoframeinterpolationgeneralized}
Shi, Z., Liu, X., Shi, K., Dai, L., Chen, J.: Video frame interpolation via
  generalized deformable convolution (2021),
  \url{https://arxiv.org/abs/2008.10680}

\bibitem{tian2023expanding}
Tian, L., Chen, F., Macosko, E.Z.: The expanding vistas of spatial
  transcriptomics. Nature Biotechnology  \textbf{41}(6),  773--782 (2023)

\bibitem{wang2024cross}
Wang, X., Huang, X., Price, S., Li, C.: Cross-modal diffusion modelling for
  super-resolved spatial transcriptomics. In: International Conference on
  Medical Image Computing and Computer-Assisted Intervention. pp. 98--108.
  Springer (2024)

\bibitem{wang2023multi}
Wang, X., Price, S., Li, C.: Multi-task learning of histology and molecular
  markers for classifying diffuse glioma. arXiv preprint arXiv:2303.14845
  (2023)

\bibitem{10.1145/3225058.3225069}
You, Y., Zhang, Z., Hsieh, C.J., Demmel, J., Keutzer, K.: Imagenet training in
  minutes. In: Proceedings of the 47th International Conference on Parallel
  Processing. ICPP '18, Association for Computing Machinery, New York, NY, USA
  (2018). \doi{10.1145/3225058.3225069},
  \url{https://doi.org/10.1145/3225058.3225069}

\bibitem{Zhang2024}
Zhang, D., Schroeder, A., Yan, H., Yang, H., Hu, J., Lee, M.Y.Y., Cho, K.S.,
  Susztak, K., Xu, G.X., Feldman, M.D., Lee, E.B., Furth, E.E., Wang, L., Li,
  M.: Inferring super-resolution tissue architecture by integrating spatial
  transcriptomics with histology. Nature Biotechnology  \textbf{42}(9),
  1372--1377 (Sep 2024). \doi{10.1038/s41587-023-02019-9},
  \url{https://doi.org/10.1038/s41587-023-02019-9}

\bibitem{zhang2024knowledge}
Zhang, Y., Wang, X., Meng, F., Tang, J., Li, C.: Knowledge-driven subspace
  fusion and gradient coordination for multi-modal learning. In: International
  Conference on Medical Image Computing and Computer-Assisted Intervention. pp.
  263--273. Springer (2024)

\bibitem{zhao2024highdimensionalbayesianmodeldiseasespecific}
Zhao, Q., Zhang, Q.: High-dimensional bayesian model for disease-specific gene
  detection in spatial transcriptomics (2024),
  \url{https://arxiv.org/abs/2409.02397}

\end{thebibliography}

\end{document}